\documentclass[aps,prc,showpacs,superscriptaddress]{revtex4}
\usepackage[latin1]{inputenc}
\usepackage{bm}
\usepackage{epsfig}
\usepackage{amsthm}
\usepackage{amsfonts}
\usepackage{float}
\usepackage{amsmath,amssymb}
\usepackage{color}
\usepackage{graphicx}
\usepackage{dcolumn}
\usepackage{bm}
\usepackage{hyperref}
\usepackage{bigints}

\def\nn {\nonumber}

\newcommand{\be}{\begin{equation}}
\newcommand{\ee}{\end{equation}}
\newcommand{\bea}{\begin{eqnarray}}
\newcommand{\eea}{\end{eqnarray}}

\newcommand{\ep}{\epsilon}
\newcommand{\om}{\omega}
\newcommand{\vk}{\vec{k}}

\newcommand{\del}{\partial}

\newcommand{\lnz}{\ln\mathcal{Z}}
\newcommand{\mt}{\left(\frac{m_h}{T}\right)}
\newcommand{\nmt}{\left(\frac{nm_h}{T}\right)}
\newcommand{\nmut}{\left(\frac{n\mu_h}{T}\right)}

\begin{document}
\title{Phenomenological bound on the viscosity of the hadron resonance gas}
\author{Snigdha Ghosh} 
\email{snigdha.physics@gmail.com}
\affiliation{Indian Institute of Technology Gandhinagar, Palaj, Gandhinagar 382355, Gujarat, India}
\affiliation{Variable Energy Cyclotron Centre, 1/AF Bidhannagar, Kolkata 700064, India}
\author{Sabyasachi Ghosh}
\email{sabyaphy@gmail.com} 
\affiliation{Indian Institute of Technology Bhilai, GEC Campus, Sejbahar, Raipur 492015, 
Chhattisgarh, India}
\author{Sumana Bhattacharyya}
\email{response2sumana91@gmail.com}
\affiliation{Center for Astroparticle Physics and Space Science, Bose Institute, 
Block EN, Sector V, Salt Lake, Kolkata 700091, India}

\begin{abstract}
We have explored some phenomenological issues during calculations of transport
coefficients for hadronic matter, produced in the experiments of heavy ion collisions.  
Here, we have used an ideal hadron resonance gas model to demonstrate the issues.
On the basis of dissipation mechanism, 
the hadronic zoo is classified into resonance and non-resonance members,
who participate in dissipation via strong decay and scattering channels respectively.
Imposing our phenomenological restriction, we are able to provide a rough upper
and lower bound estimations of transport coefficients. Interestingly, we find that our 
proposed lower limit estimation for shear viscosity to entropy density ratio is
little larger than its quantum lower bound. By taking a simple example,
we have demonstrated how our proposed restriction help to tune any estimation
of transport coefficients within its numerical band, proposed by us. 
\end{abstract}

\maketitle

\section{Introduction} 
Shear viscosity $\left(\eta\right)$ to entropy density $(s)$ ratio is the measure of fluidity of the medium. 
Being roughly proportional to the ratio of mean free path to de-Broglie wavelength of 
medium constituent, the $\eta/s$ of any fluid can never be vanished, because mean free path 
of any constituent can never be lower than its de-Broglie wavelength. It indicates that 
quantum fluctuations prevent the existence of perfect fluid in nature and $\eta/s$ of 
any fluid should have some lower bound, which is also claimed from the 
string theory calculation~\cite{KSS}. 
Interestingly, a small value of $\eta/s$, close to this quantum lower bound, is observed
in super hot medium, produced in Relativistic Heavy Ion Collider (RHIC) experiment
as well as in some other
many body systems like cold atoms~\cite{coldatom}, graphene~\cite{graphene} and in
low energy nuclear matter~\cite{VECC}. This nearly perfect fluid behavior, at extreme conditions, has drawn 
immense attention from scientific communities working on the field of 
condense matter physics to nuclear physics to string theory.

In our present work, we emphasize on some phenomenological issues of $\eta/s$ for hadronic matter.
We get a long list of Refs.~\cite{Prakash,Itakura,Dobado,Nicola,Weise1,SSS,Ghosh_piN,
Gorenstein,HM,Hostler,Purnendu,Redlich_NPA,Marty,G_CAPSS,G_IFT,Deb,Tawfik,Bass,Muronga,Plumari,Pal},
which had addressed different microscopic calculations of this $\eta/s$,
based on different hadronic 
models~\cite{Prakash,Itakura,Dobado,Nicola,Weise1,SSS,Ghosh_piN,Gorenstein,HM,Hostler}, different 
effective QCD models~\cite{Purnendu,Redlich_NPA,Marty,G_CAPSS,G_IFT,Deb,Tawfik} and bulk 
simulations~\cite{Bass,Muronga,Plumari,Pal}. 
The predicted values of $\eta/s$ from earlier estimations
reside within a broad numerical band. Same is observed for bulk viscosity 
$\zeta$~~\cite{{Nicola},{HM},{Hostler},{Purnendu},{Redlich_NPA},{Marty},{G_IFT},{Deb},{Tawfik},{Prakash},{Gavin},
{Paech1},{Arnold_bulk},{Santosh},{Meyer_zeta},{Dobado_zeta1},{Dobado_zeta2},{Redlich_PRC},{De-Fu},{Tuchin},{Tuchin2},
{Vinod},{Nicola_PRL},{Sarkar},{SG_NISER},{Sarwar},{Kadam10},{Kinkar_bulk}}.

Here we have found a possibility of comparatively narrower band for $\eta/s$
of hadronic matter, when we put a restriction in the calculations
of $\eta/s$, based on an ideal hadron resonance gas (HRG) model.
The restriction is to consider the dissipation of hadrons within a finite
size of RHIC or LHC matter. When we follow the expressions of different transport coefficients in
the framework of relaxation time approximation (RTA), we assume that relaxation length (time) should
be lower than the size (life time) of the system or medium. Owing to the fact, when we consider the hadronic matter,
the resonances, whose mean life time are larger than the life time of the system, will not take part
in dissipation process. So we have to eliminate them during the calculation of transport coefficients for finite
size hadronic matter. On the other hand, the hadrons like pion, kaon, nucleon can have a momentum dependent
relxation length, whose high momentum component may become larger than the system size. Therefore, we have
to eliminate the high-momentum part by imposing a upper momentum cut-off in the calculation.
This fact of finite size dissipation is pointed out in the present work with the help of ideal HRG model.
A generic qualitative message of the present study is that the theoretical tools should have to
take care of this fact of finite size dissipation when we try to give the estimation of transport
coefficients for RHIC or LHC matter.

The article is organized as follows. 
Next, in the formalism part (\ref{Form}), first (\ref{Form1}) we have addressed the standard expression
of different transport coefficients and then (\ref{Form2}) we have provided a breif description
of ideal HRG model, whose detail expressions are given in the Appendix (\ref{appendix.hrg}). 
After getting the expression of transport coefficients and thermodynamical 
quantity like entropy density, they have been folded by spectral function of hadrons. Its generic
equation is written in the subsection (\ref{Form3}) of formalism part. Next we will come to the 
results section (\ref{Res}), where we have explored the issues of finite size dissipation,
which can give us a rough numerical band in the values of different transport coefficients.
Then, we have provided an example of microscopic calculation
of transport coefficients, whose values don't remain within our proposed band
but after utilizing the appropriate restriction of finite size dissipation, we get their modified
values, which ultimately remain within our proposed band. Atlast, we summarize our studies in section (\ref{Sum}). 

\section{Formalism}
\label{Form}
\subsection{Transport Coefficients in Kubo Formalism}
\label{Form1}
Our aim of this work is to calculate these transport coefficients with the 
help of HRG model, so we have to add the contributions 
of all mesons ($M$) and baryons ($B$) for getting total transport coefficients
of hadronic matter. We know that the mathematical structure of transport coefficients,
obtained from the one-loop diagram in quasi-particle Kubo approach and 
relaxation time approximation (RTA) in kinetic theory approach, are exactly same.
Without going those background formalism part of transport coefficients like shear viscosity
$\eta$~\cite{Nicola,Weise1,Gavin,G_IJMPA} and bulk viscosity $\zeta$~\cite{Nicola,G_IFT}
therefore, let us start with their standard expressions:
%
\bea
\eta&=&\sum_{h\in\{\text{hadrons}\}}\frac{g_h}{15 T}\bigintsss \frac{d^3\vk}{(2\pi)^3}
\tau_{h}\left(\frac{\vk^2}{\om_{h}}\right)^2\frac{}{}f_h\left(1-a_hf_h\right)~, 
\label{eta_G} \\
\zeta&=&\sum_{h\in\{\text{hadrons}\}}\frac{g_h}{T}\bigintsss \frac{d^3\vk}{(2\pi)^3\om_{h}^2}
\tau_{h}\left\{\left(\frac{1}{3}-c_s^2\right)\vk^2  - c_s^2 m_h^2\right\}^2
f_h\left(1-a_hf_h\right)~,
\label{zeta_G} 
\eea

where $g_{h}$, $\om_h=\sqrt{\vk^2 +m_h^2}$ and $f_h = \left[ e^{\om_h/T} +a_h \right]^{-1}$ are 
respectively the degeneracy factor, energy and thermal distribution function 
(Fermi-Dirac or Bose-Einstein) of hadron $h$;
 $a_h = \pm 1$ if $h$ is a Fermion/Boson.
In the above equation, $\tau_{h}$ is
the relaxation time of $h$ which proportionally controls the numerical
strength of the transport coefficients. Obviously, the thermal phase space factors, depend on the 
thermal distribution functions of different hadrons, are another controlling
component for transport coefficients. 

\subsection{Thermodynamics from ideal HRG }
\label{Form2}

As we are interested on the (nearly) perfect fluid nature of the medium,
produced in HIC experiments, so we focus on the quantity - fluidity, 
which is quantified by the $\eta/s$, where $S$ is entropy density. 
To calculate $s$ of hadronic matter, we follow the standard
procedure of ideal HRG model~\cite{HRGrev}, where all thermodynamic quantities like
energy density ($\ep$), pressure ($P$), entropy density ($s$), speed of sound ($c_s$) etc. 
are calculated from the partition function. 
The Grand Canonical partition function is given by , 
\begin{equation}
\ln\mathcal{Z}\left(T,V,\{\mu\}\right) = V\bigintss\frac{d^3p}{(2\pi)^3}\sum_{h\in\{\text{hadrons}\}}^{}g_ha_h\ln\left[\frac{}{}1+a_h
\exp\left\{-\beta\left(\omega_h-\sum_{\mu_k\in\{\mu\}}^{}q_h^k\mu_k\right)\right\}\right]
\label{eq.partition.function}
\end{equation}
where $\{\mu\} = \left\{ \mu_B,\mu_Q,\mu_S,....\right\}$ is the set of chemical potentials corresponding to 
the conserved quantities (like net baryon ($n_B$), net charge ($n_Q$), net strangeness ($n_S$) etc. ) and $q_h^k$ is 
the corresponding quantum number of $h^\text{th}$ hadron. 
From the partition function, all the thermodynamic quantities can be calculated:
\begin{eqnarray}
P &=& \left(\frac{T}{V}\right)\lnz 
\label{eq.pressure}\\
\varepsilon &=& \left(\frac{T^2}{V}\right)\frac{\del}{\del T}\left(\lnz\right) \label{eq.energy}\\
n_k &=& \left(\frac{T}{V}\right)\frac{\del}{\del\mu_k}\left(\lnz\right) ~~;~~ k = B,Q,S,.... 
\label{eq:nk} \\
 c_s^2 &=& \left(\frac{\partial p}{\partial\epsilon}\right) 
 = \left(\frac{\partial p}{\partial T}\right)\Big/\left(\frac{\partial\epsilon}{\partial T}\right)
 +\sum_{\mu_k\in\{\mu\}}^{}\left(\frac{\partial p}{\partial\mu_k}\right)\Big/\left(\frac{\partial\epsilon}{\partial\mu_k}\right)~.
\end{eqnarray} 
The entropy density $s$ can be obtained from
\begin{eqnarray}
s = \left(\frac{\varepsilon+P}{T}\right)-\frac{1}{T}\sum_{\mu_k\in\{\mu\}}^{}n_k\mu_k~. \label{eq.entropy}
\end{eqnarray}
The momentum integration in Eq.~(\ref{eq.partition.function}) can be analytically performed in terms of modified 
Bessel function details of which are provided in Appendix~\ref{appendix.hrg}. In this work we have taken  
$\mu_B=\mu_Q=\mu_S=....=0$ which implies that $\{\mu\}$ is a null set.

\subsection{Spectral Folding}
\label{Form3}

The transport coefficients as well as the thermodynamic quantities as given in Eqs~(\ref{eta_G})-(\ref{zeta_G}), 
(\ref{eq.pressure})-(\ref{eq.entropy}), depend on the masses of all the hadrons $\{m_h\}$. 
To take into account the finite widths of the unstable hadrons, the transport coefficients and the 
thermodynamic quantities are folded with the corresponding hadronic spectral functions $\rho_h^\text{m,b}(M)$. 
Let $\Phi$ denotes any of the transport coefficients (such as $\eta$, $\zeta$) or thermodynamic quantities 
(such as $\varepsilon$, $P$, $s$ etc.). In this work, the spectral foldings are done through, 
\begin{eqnarray}
\Phi_\text{folded} = \sum_{h\in\{\text{mesons}\}}^{}\frac{1}{N_h^\text{m}}\int^{(m_h+2\Gamma_h)^2}_{(m_h-2\Gamma_h)^2}dM^2 
\rho_h^\text{m}(M) \Phi\left(m_h=M\right)  \nn \\ 
+ \sum_{h\in\{\text{baryons}\}}^{} \frac{1}{N_h^\text{b}}\int^{m_h+2\Gamma_h}_{m_h-2\Gamma_h}dM
\rho_h^\text{b}(M) \Phi\left(m_h=M\right)~,
\label{fold}
\end{eqnarray}
with $N_h^\text{m}=\displaystyle\int^{(m_h+2\Gamma_h)^2}_{(m_h-2\Gamma_h)^2}dM^2\rho_\text{m}(M)$  and 
 $N_h^\text{b}=\displaystyle\int^{m_h+2\Gamma_h}_{m_h-2\Gamma_h}dM \rho_\text{b}(M)$.  
In the above equation, the mesonic and baryonic spectral functions are respectively,
\begin{eqnarray}
\rho_h^\text{m}(M) &=& \frac{1}{\pi} {\rm Im} \left[\frac{1}{M^2-m_h^2 + iM\Gamma_h } \right] \\
\rho_h^\text{b}(M) &=& \frac{1}{\pi} {\rm Im} \left[\frac{1}{M-m_h + \frac{i}{2}\Gamma_h} \right]~.
\end{eqnarray}

%
%
\section{Numerical Results}
\label{Res}

\begin{figure}[h]
	\includegraphics[scale=0.23,angle=-90]{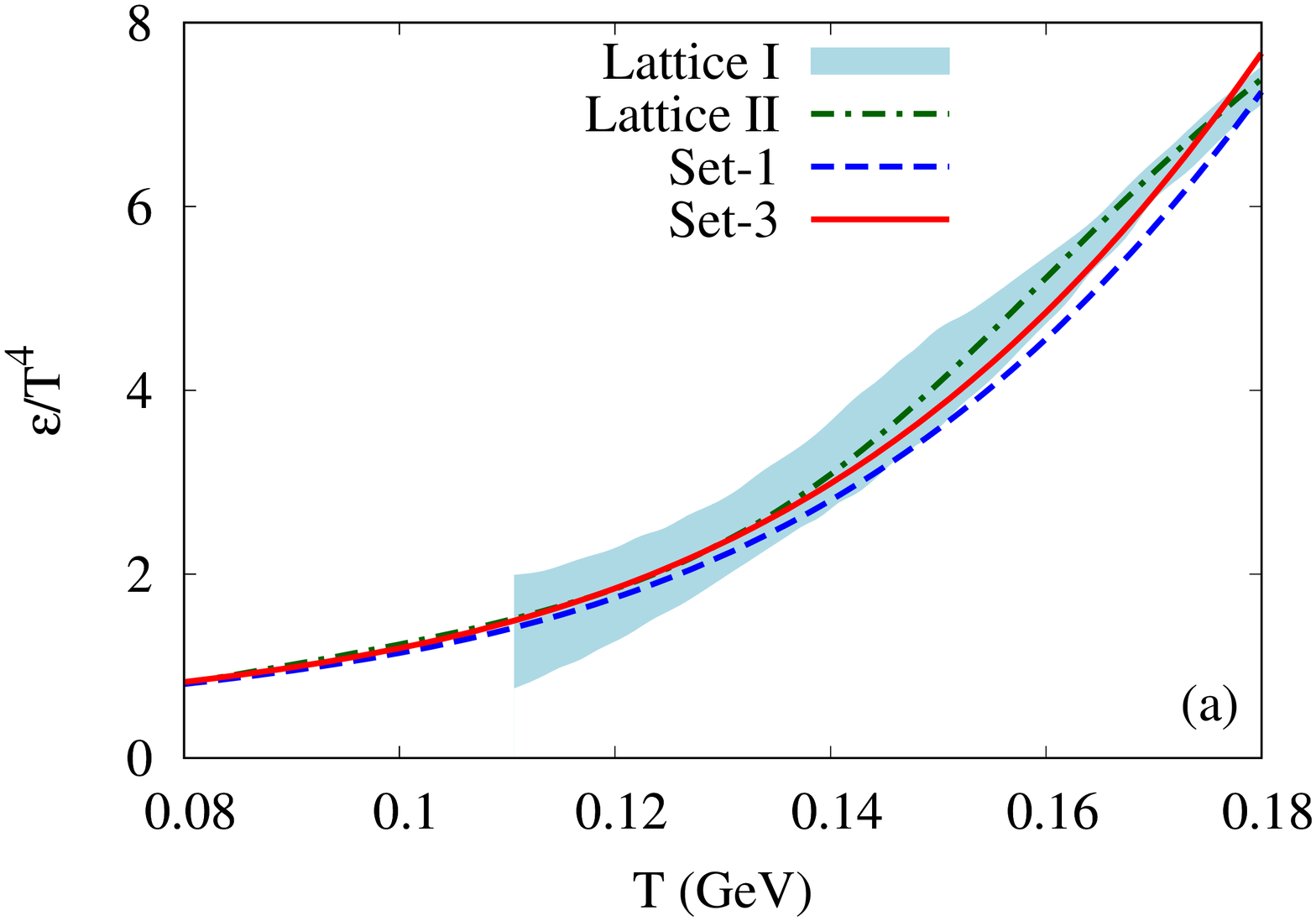}
	\includegraphics[scale=0.23,angle=-90]{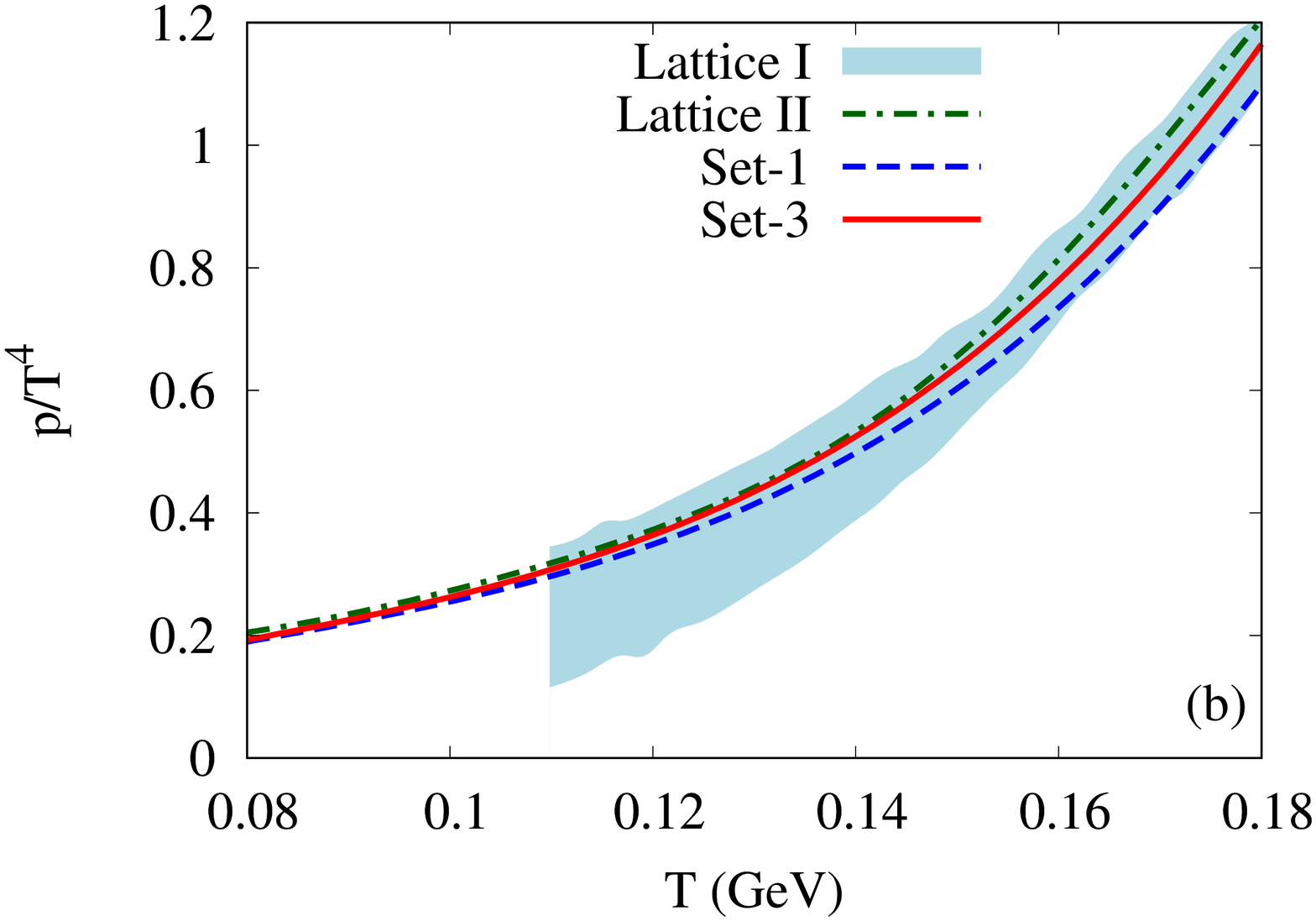}
	\includegraphics[scale=0.23,angle=-90]{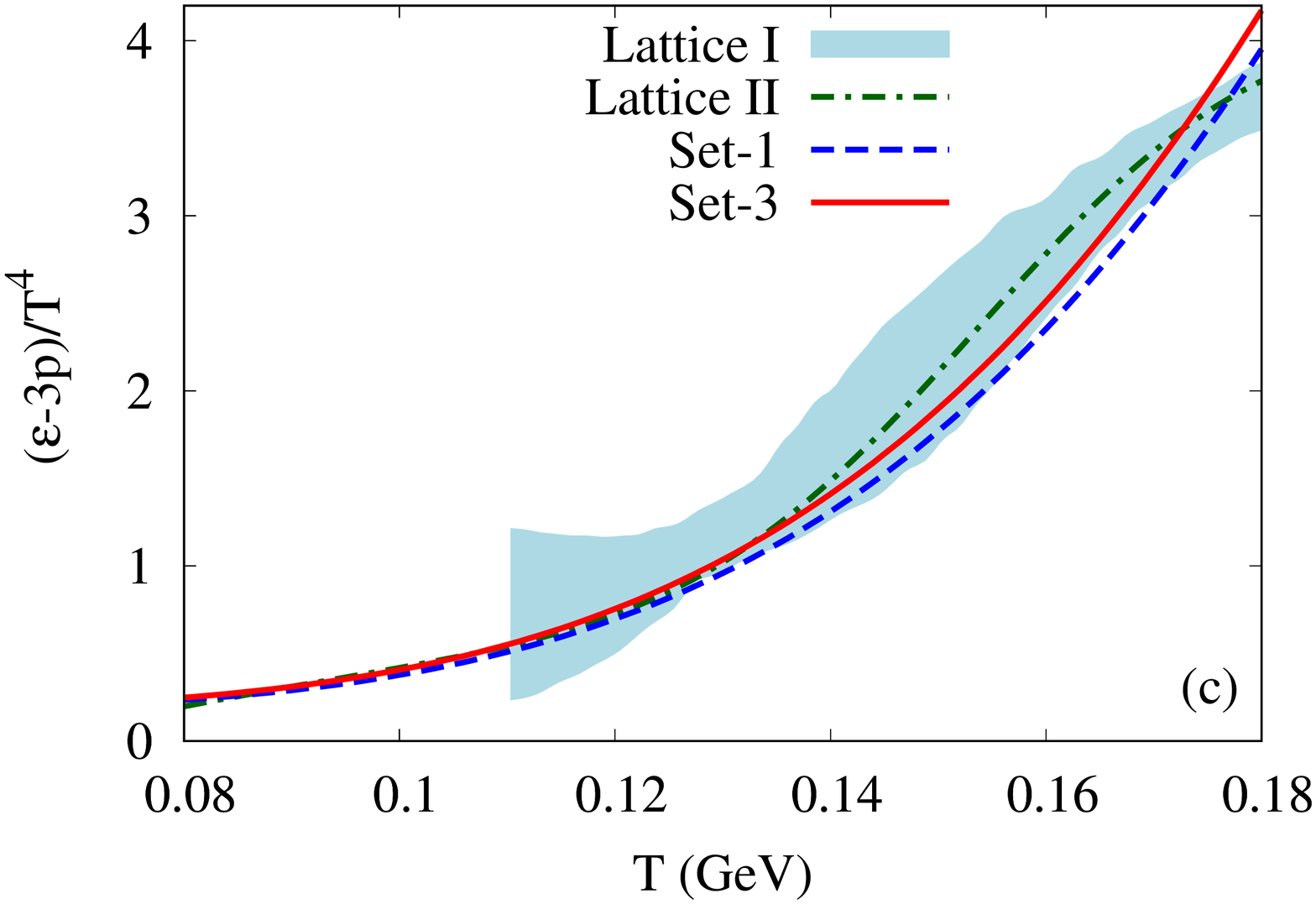}
	\caption{(Color online) (a) Energy density (b) Pressure and (c) Trace anomaly 
		scaled with fourth power of inverse temperature as a function of temperature  
		compared among results from ideal HRG and two lattice QCD data from Ref.~\cite{Borsanyi:2013bia} and 
		\cite{LQCD_2014} abbreviated as Lattice I and Lattice II respectively. Set-1 and Set-3 (see Table~\ref{tab2}) corresponds to results from 
		ideal HRG without and with spectral folding.}
	\label{fig:eptrace} 
\end{figure}

\begin{figure}[h]
	\includegraphics[scale=0.3,angle=-90]{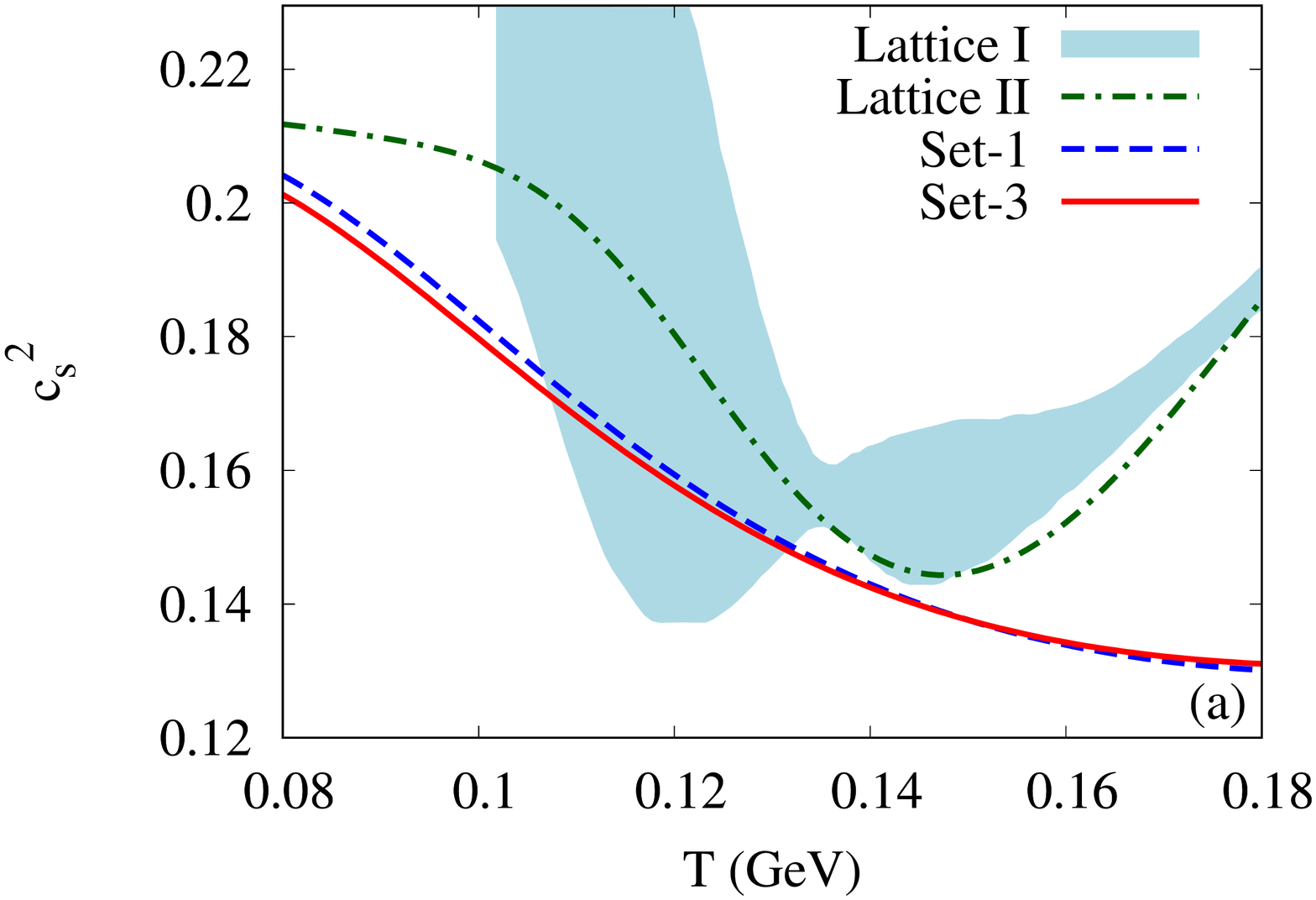}
	\includegraphics[scale=0.3,angle=-90]{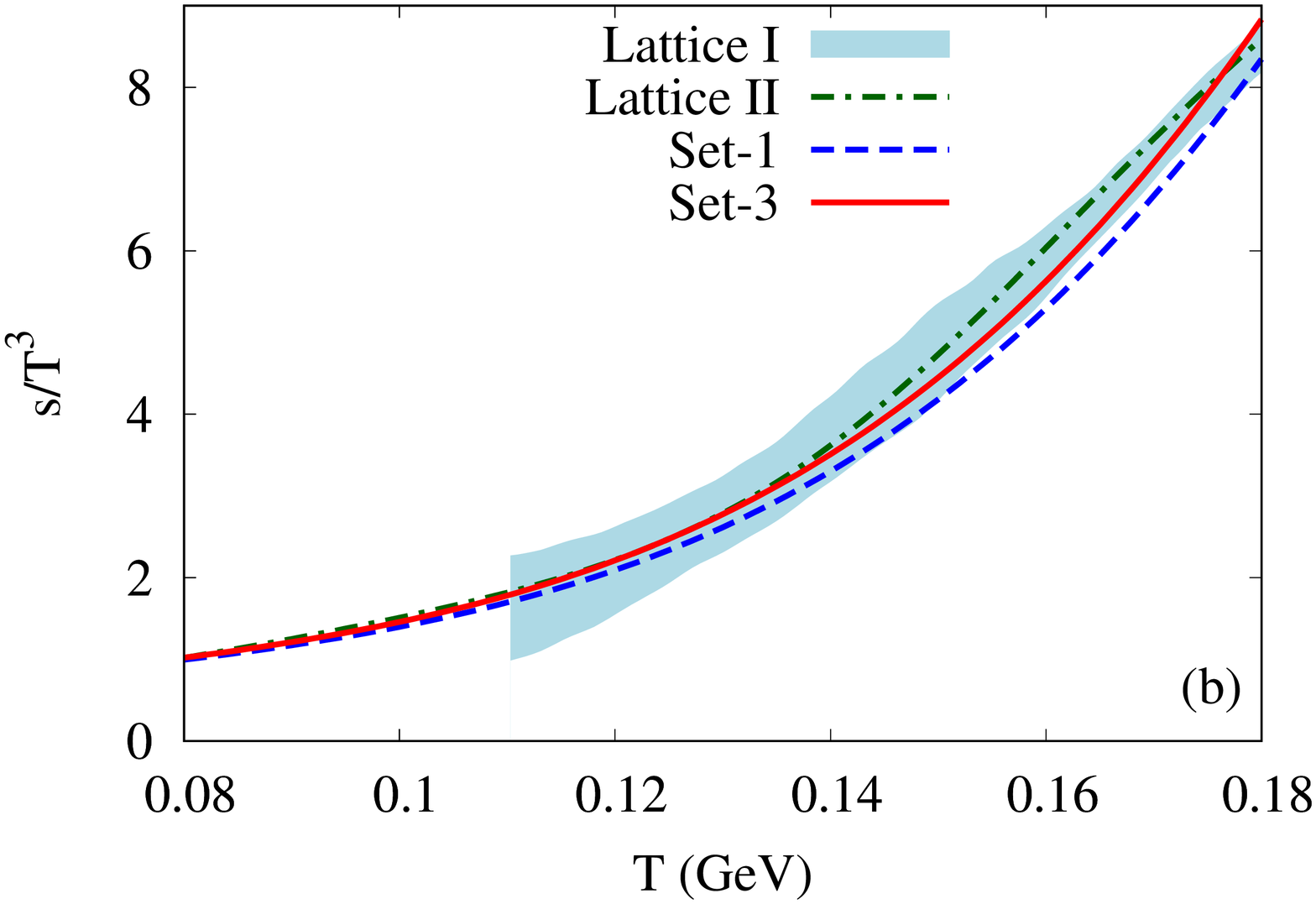}
	\caption{(Color online) (a) Speed of sound (b) Entropy density scaled with third power of inverse temperature as a 
		function of temperature compared among results from ideal HRG and two lattice QCD data from Ref.~\cite{Borsanyi:2013bia} and \cite{LQCD_2014}  
		abbreviated as Lattice I and Lattice II respectively. Set-1 and Set-3 (see Table~\ref{tab2}) corresponds to results from 
		ideal HRG without and with spectral folding.}
	\label{fig:cs2entropy} 
\end{figure}

Let us begin this section by showing numerical results for the thermodynamic quantities 
obtained from ideal HRG model in Figs~\ref{fig:eptrace} and \ref{fig:cs2entropy}. 
From Eqs.~(\ref{eq.energy}) and (\ref{eq.pressure}), one can obtain $\ep$ and $P$,
which are shown by blue dash line in Fig.~\ref{fig:eptrace}-(a), (b). Using
the folding technique, as given in Eq.~(\ref{fold}),
the values of $\ep$ and $P$ are little bit enhanced as shown by red line in
Fig.~\ref{fig:eptrace}-(a), (b). Similar kind of results for 
trace anomaly $(\ep - 3P)/T^4$ is shown in Fig.~\ref{fig:eptrace}-(c).
One of the success of HRG model is that its estimated values of different 
thermodynamical quantities are quite close to the results,
obtained by Lattice Quantum Chromo Dynamics (LQCD). We have added
two set of lattice QCD data from Ref.\cite{Borsanyi:2013bia} (cyan band) and \cite{LQCD_2014} (green dash dot line),
which are well agreement with the HRG results of present work.
In same pattern, results of $C_S^2$ and entropy density $s$ are also
plotted in Fig.~\ref{fig:cs2entropy}(a) and (b).

Now let us come to the results of transport coefficients.
In the expression of $\eta$, $\zeta$, given in Eqs.~(\ref{eta_G}), (\ref{zeta_G}), we
see that the thermodynamical phase space parts of different hadrons
are known components from the HRG model but their relaxation times
are unknown components, which we should have to include in the model 
from outside, based on our phenomenological understanding. Owing to
this phenomenological picture of relaxation of different hadrons,
we have classified them into two categories - non-resonance (NR) and
resonance (R) components.  Let us call
pseudo-scalar meson nonet and baryon octet as NR members
as these long lived particles can't decay inside the fireball, 
produced in HIC experiments. 
Among them, pion, kaon and nucleon are most abundant
constituents in the medium, hence we consider only them as NR
members for simplicity. Their strong interaction elastic scattering
will provide their relaxation times, which are expected to be important
in dissipation within the life time of fireball. 
The hadrons, other than pseudo-scalar meson nonet and baryon octet,
are considered as R members as maximum of them follow
strong decays. Their mean life times which are the inverse of their strong decay widths, 
are comparable with the life time of fireball.
So these hadrons also live in the fireball 
along with the NR members but they exist in resonance states.
Maximum of them decays within the medium during its life time and
therefore, their strong decays contributes to the total dissipation
of the medium. So, by ignoring other interactions (weak and electromagnetic 
decays and scattering channels)~\cite{Purnendu}, 
the strong interaction scale is our matter of interest
for calculating dissipation in the medium, which survives in that scale.
So we see that R and NR members participate in dissipation
by their strong decay and scattering processes respectively.
\begin{figure}[h]
\includegraphics[scale=0.3,angle=-90]{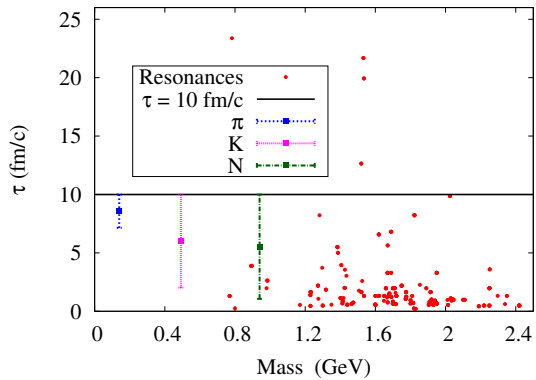}
\caption{(Color online) The values of mean life times (red points) for 
different hadron {\it resonances} up to $2.5$ GeV masses. 
The horizontal line indicates an approximated 
life time of the hadronic medium, produced in heavy ion experiments.
Blue, pink and green bars are denoting the ranges of collisional time
or relaxation time for {\it non-resonance} members $\pi$, $K$ and $N$ respectively.
}
\label{particles_2} 
\end{figure}
\begin{figure}[h]
\includegraphics[scale=0.3,angle=-90]{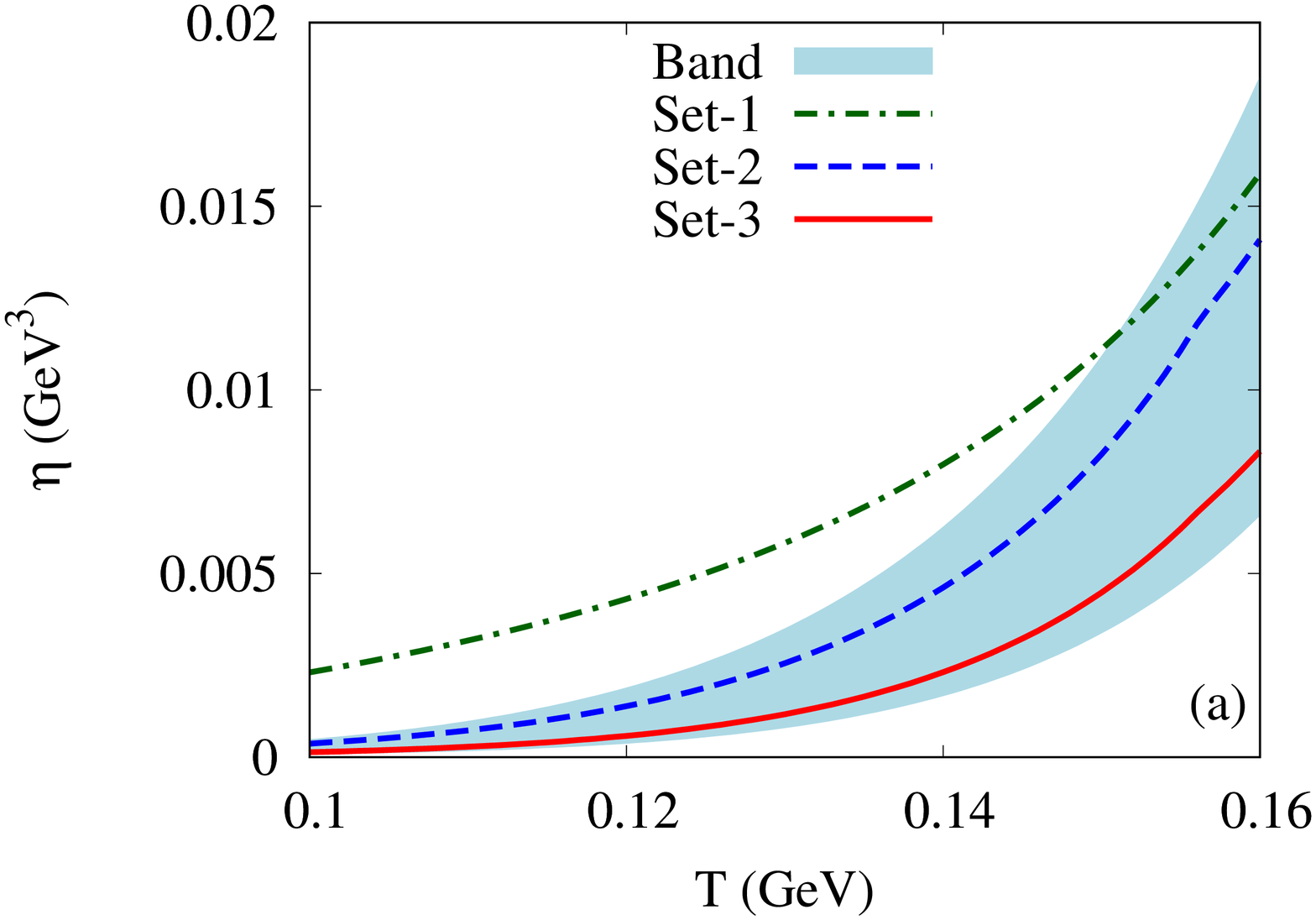}
\includegraphics[scale=0.3,angle=-90]{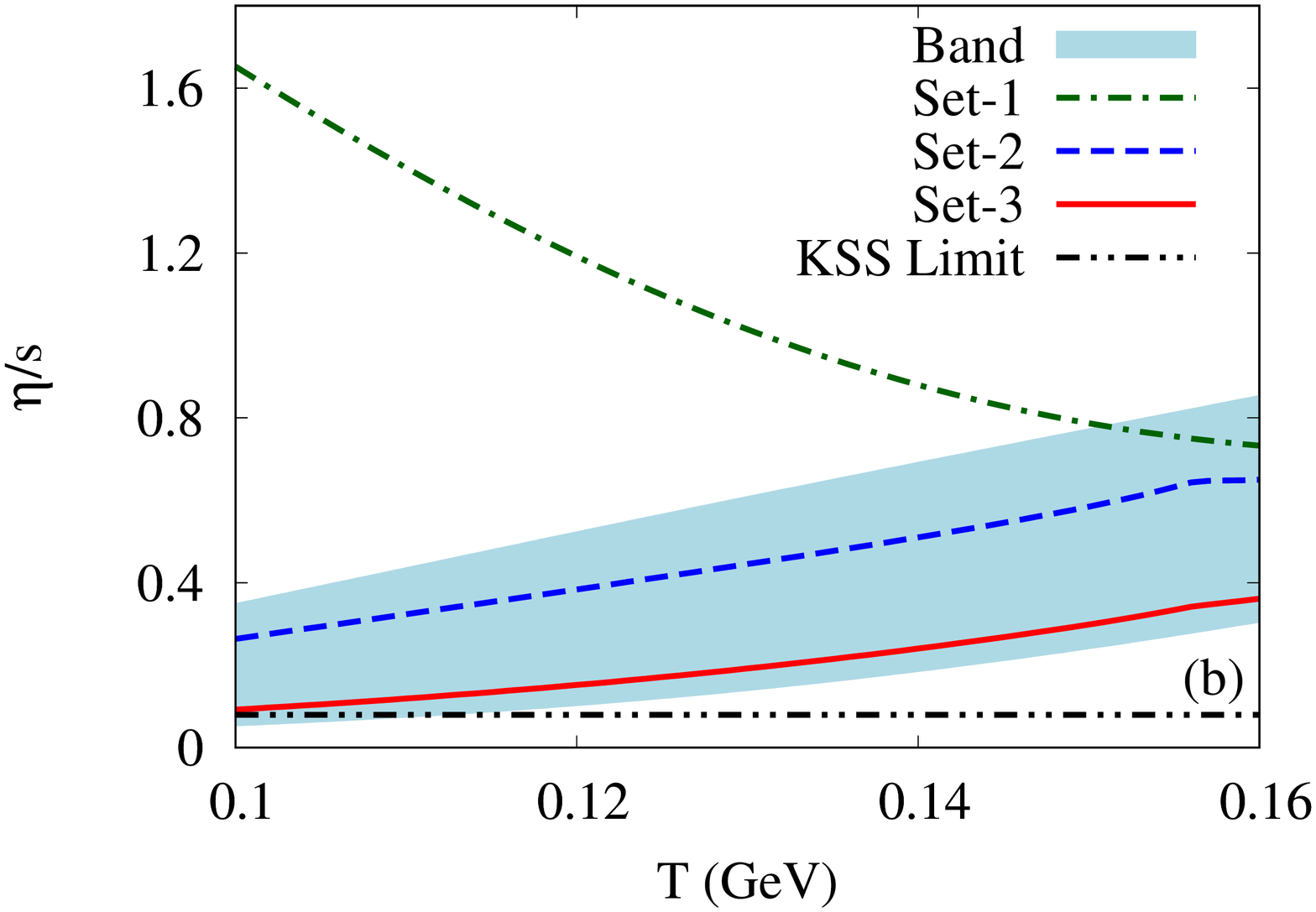}
\caption{(Color online) (a) Shear viscosity ($\eta$) (b) shear viscosity to entropy density ratio ($\eta/s$) 
	as a function of temperature for Set-1, Set-2 and Set-3, as given in Table~\ref{tab2}. 
	An approximate numerical band of both are shown by cyan color and the dash-dot-dot line indicates 
	the KSS limit for $\eta/s$.}
\label{fig:eta} 
\end{figure}
\begin{figure}[h]
	\includegraphics[scale=0.3,angle=-90]{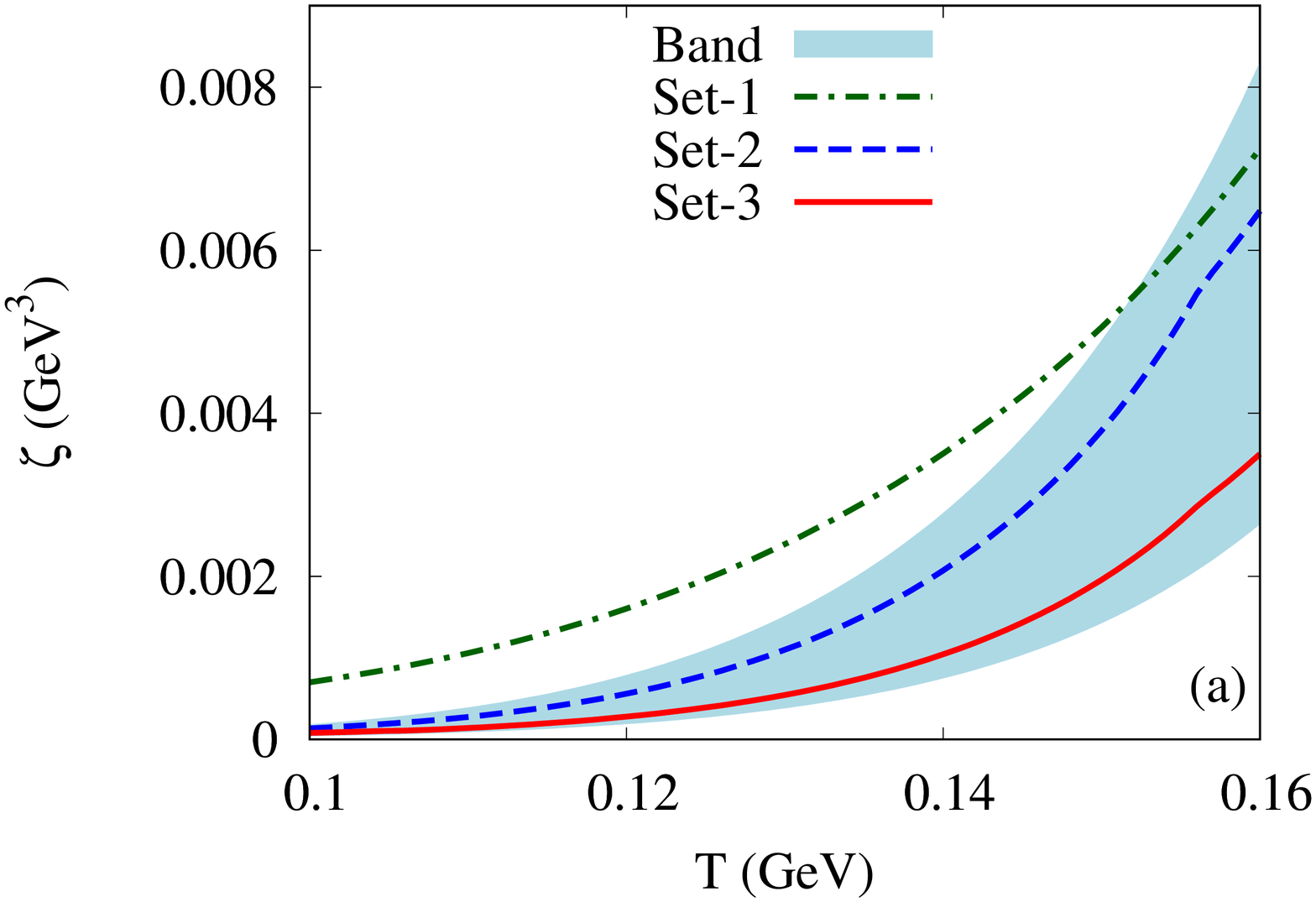}
	\includegraphics[scale=0.3,angle=-90]{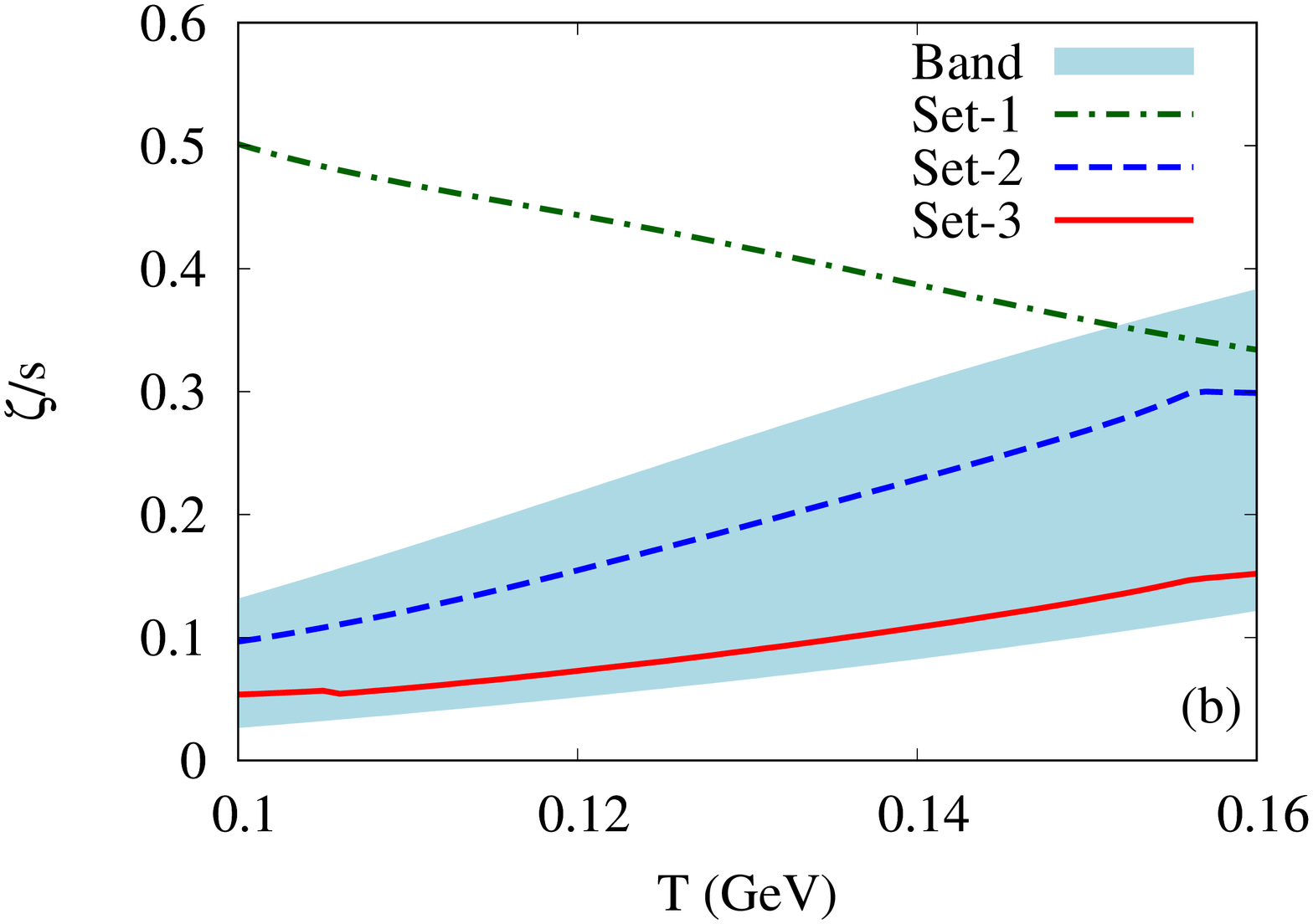}
	\caption{(Color online) The temperature dependence of (a) Bulk viscosity ($\zeta$) 
	(b) bulk viscosity to entropy density ratio ($\zeta/s$)
		as a function of temperature for Set-1, Set-2 and Set-3 as given in Table~\ref{tab2}. 
		Their approximate numerical band is also shown.}
	\label{fig:zeta} 
\end{figure}
%

%

To explore our phenomenological studies on the dissipation process, we have plotted
mean life times (red circles) of different hadrons upto $M=2.5$ GeV in 
Fig.~\ref{particles_2}, which is basically covering the strong interaction spectra of hadronic zoo
at a glance. The horizontal blue dash line indicates the life time of fireball,
which is approximately taken as $10$ fm. 
Hence, the hadrons, whose mean life times are less than the life time of medium,
decay inside the medium and they will only participate in the dissipation.
Considering only those hadrons and using their mean life times as the relaxation 
times $\tau$ in Eq.~(\ref{eta_G}), we will get shear viscosity of R component.
One has to always consider this amount of shear viscosity for hadronic matter,
which may be considered as a lower estimation of $\eta$ in HRG even when
we don't take any NR contribution. Now, we focus on the NR
component, whose in-medium scattering contribution will not be a fixed value like
R component. In different hadronic model 
calculation~\cite{Prakash,Itakura,Dobado,Nicola,Weise1,SSS,Ghosh_piN}, we
notice different numerical strength of $\eta$ from this NR component,
although some refs.~\cite{Dobado,Nicola,Weise1,SSS} are concentrated only in pion medium.
Let us take Compton lengths $(1/m_{\pi,K,N})$ of NR particles ($\pi$, $K$, $N$)
as their minimum scattering lengths in Eq.~(\ref{eta_G}) and then add this contribution
with R component to get an approximate lowest estimation of total $\eta$.
On the other hand, life time (maximum size) of fireball can be considered as upper
limit of relaxation times (relaxation lengths) of NR particles
and after adding this contribution with R part contribution, we
get an upper limit estimation of $\eta$. These ranges of relaxation times for $\pi$,
$K$ and $N$ are shown by blue, pink and green bars in Fig.~\ref{particles_2}
and using these ranges, we get a numerical band of $\eta$, which is
shown by cayan color in Fig.~\ref{fig:eta}(a).

Normalizing this $\eta$ by that entropy density $s$, we get
a similar numerical band for $\eta/s$ ratio, as shown in Fig.~\ref{fig:eta}(b).
%
%
In Fig.~\ref{fig:eta}(b), we show that lowest possible value of $\eta/s$
is little greater than its quantum lower bound ($\frac{\eta}{s}=\frac{1}{4\pi}\approx 0.08$).
%
At $T=0.160$ GeV, our proposed band provides an approximate
inequality $0.3<\frac{\eta}{s}<0.85$. Now, analyzing the earlier estimations of 
$\eta/s$ for hadronic matter~\cite{Itakura,Nicola,Weise1,Ghosh_piN,
Gorenstein,HM,Hostler,Bass,Pal}, we see that 
$\eta/s(T=0.160 {\rm GeV})\approx 0.8$~\cite{Itakura}, $0.45$~\cite{Gorenstein},
$0.32$~\cite{Weise1}, $0.3$~\cite{Nicola,Hostler} remain within the inequality,
except $\eta/s(T=0.160 {\rm GeV})\approx 1$~\cite{Bass,Pal}, $0.2$~\cite{HM} 
and $0.13$~\cite{Ghosh_piN}. 
Whereas, at freeze out temperature $T=0.100$ GeV (say), their $\eta/s$ 
($\approx 2$~\cite{Gorenstein}, $1.2$~\cite{Itakura}, $1$~\cite{HM,Bass, Pal},
$0.9$~\cite{Nicola,Weise1}, $0.45$~\cite{Ghosh_piN}, $0.4$~\cite{Hostler}) are
not at all located within our proposed inequality $0.007<\eta/s<0.4$. Absence
of R members in some formalism~\cite{Itakura,Nicola,Weise1,Ghosh_piN}
and absence of considering dissipation of hadrons within the finite size hadronic matter in 
Refs.~\cite{Gorenstein,HM,Hostler,Bass,Pal} may be possible reason
for being outside of our proposed band. 

Similar kind of numerical band can be
obtained from standard RTA expressions of 
bulk viscosity ($\zeta$) as given in
Eq.~(\ref{zeta_G}).
%
%
These bands are shown by cyan color 
in Fig.~(\ref{fig:zeta}). 
%
%
%
\begin{table}[h]
\caption{Experimental values of scattering lengths of two body
elastic scattering of $\pi$, $N$ and $K$, taken from 
Refs.~\cite{exp1,exp2,exp3,exp4,exp5}.}
\begin{tabular}{|l|c|c|}
\hline 
~ & ~ & ~ \\
$HH$ & $a_{HH}^I$ & $\sigma_{HH}=\frac{\sum (2I+1)4\pi |a_{HH}^I|^2}{\sum (2I+1)}$ \\
~ & ~ & ~ \\
\hline
& $a_{\pi\pi}^{I=0}=+0.37$ fm &  \\
$\pi\pi$      &     & $\sigma_{\pi\pi}=17.3$ mb     \\
& $a_{\pi\pi}^{I=0}=-0.04$ fm &  \\
\hline
& $a_{\pi N}^{I=1/2}=+0.24$ fm & \\
$\pi N$   &    &  $\sigma_{\pi N}=16.1$ mb    \\
& $a_{\pi N}^{I=3/2}=-0.14$ fm &   \\
\hline
& $a_{NN}^{I=0}=+20.1$ fm &  \\
$NN$ &    &   $\sigma_{N N}=53.4$ b     \\
& $a_{NN}^{I=1}=-5.4$ fm &  \\
\hline
& $a_{K N}^{I=0}=-0.007$ fm &   \\
$KN$ &   &  $\sigma_{K N}=4.7$ mb  \\
& $a_{K N}^{I=1}=-0.225$ fm &   \\
\hline
& $a_{K\pi}^{I=1/2}=-0.22$ fm &   \\
$K\pi$ &  &  $\sigma_{K N}=12.2$ mb   \\
& $a_{K\pi}^{I=3/2}=-0.04$ fm &   \\
\hline
\end{tabular}
\label{tab1}
\end{table}
\begin{figure}[h]
\includegraphics[scale=0.23,angle=-90]{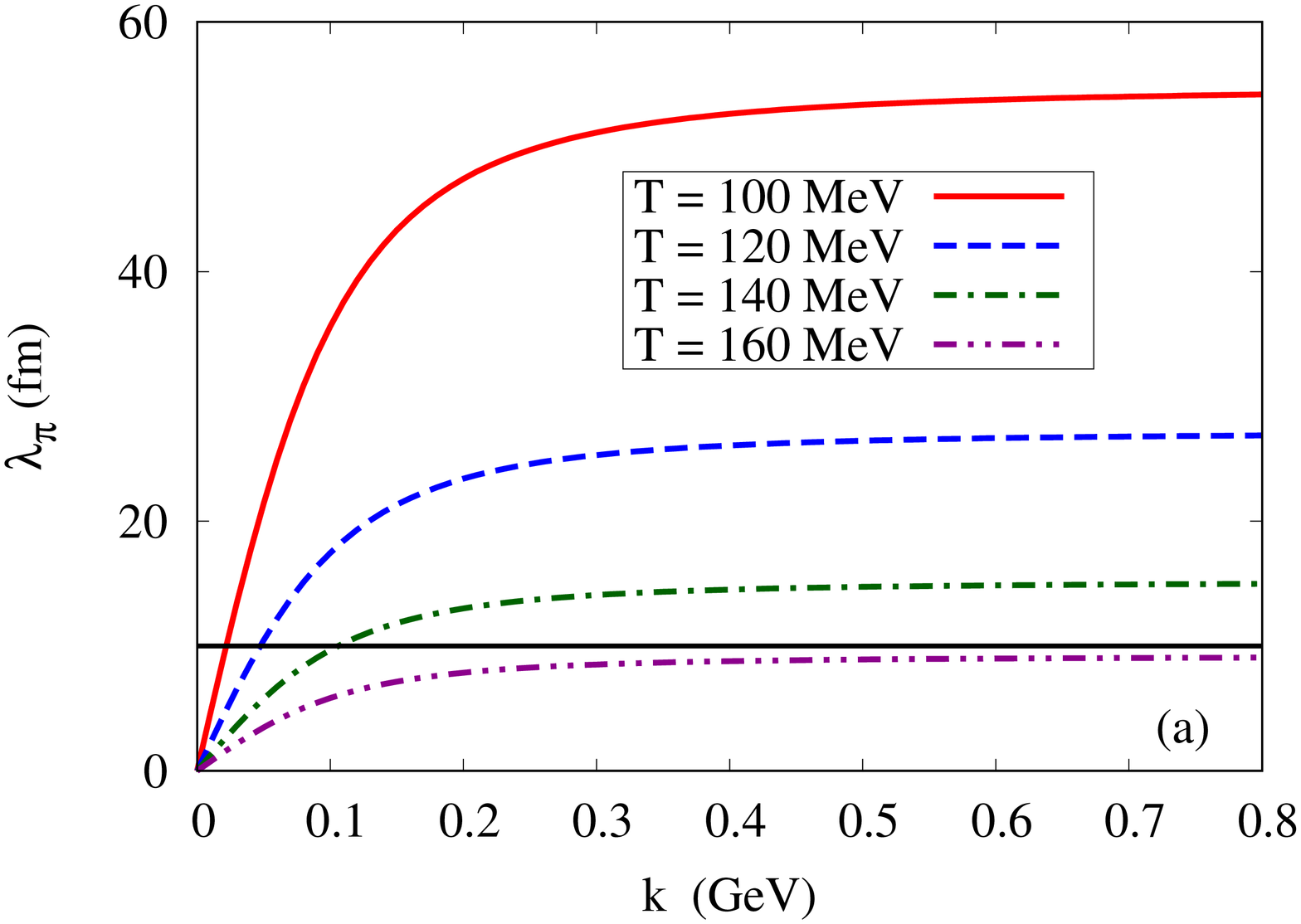}
\includegraphics[scale=0.23,angle=-90]{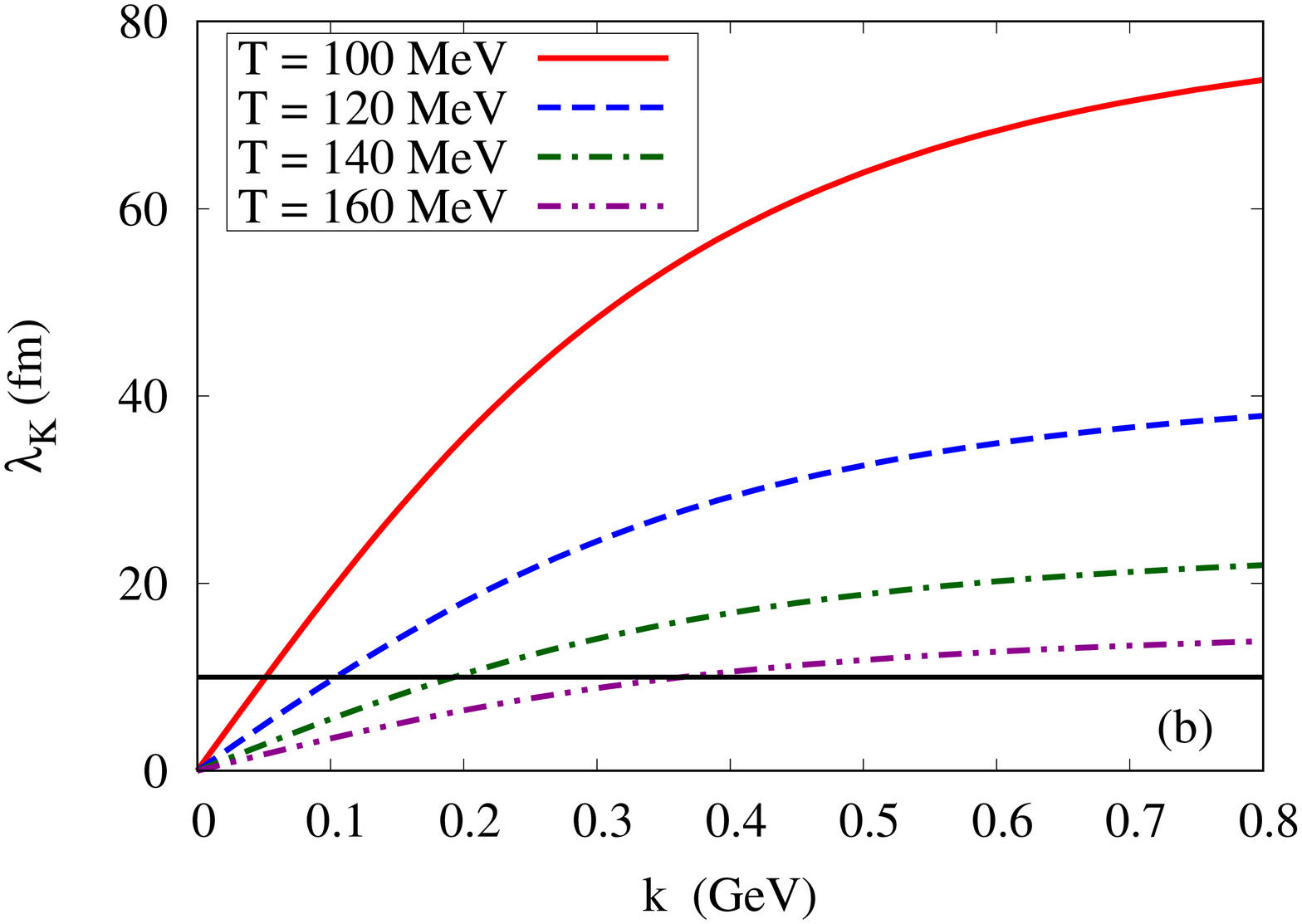}
\includegraphics[scale=0.23,angle=-90]{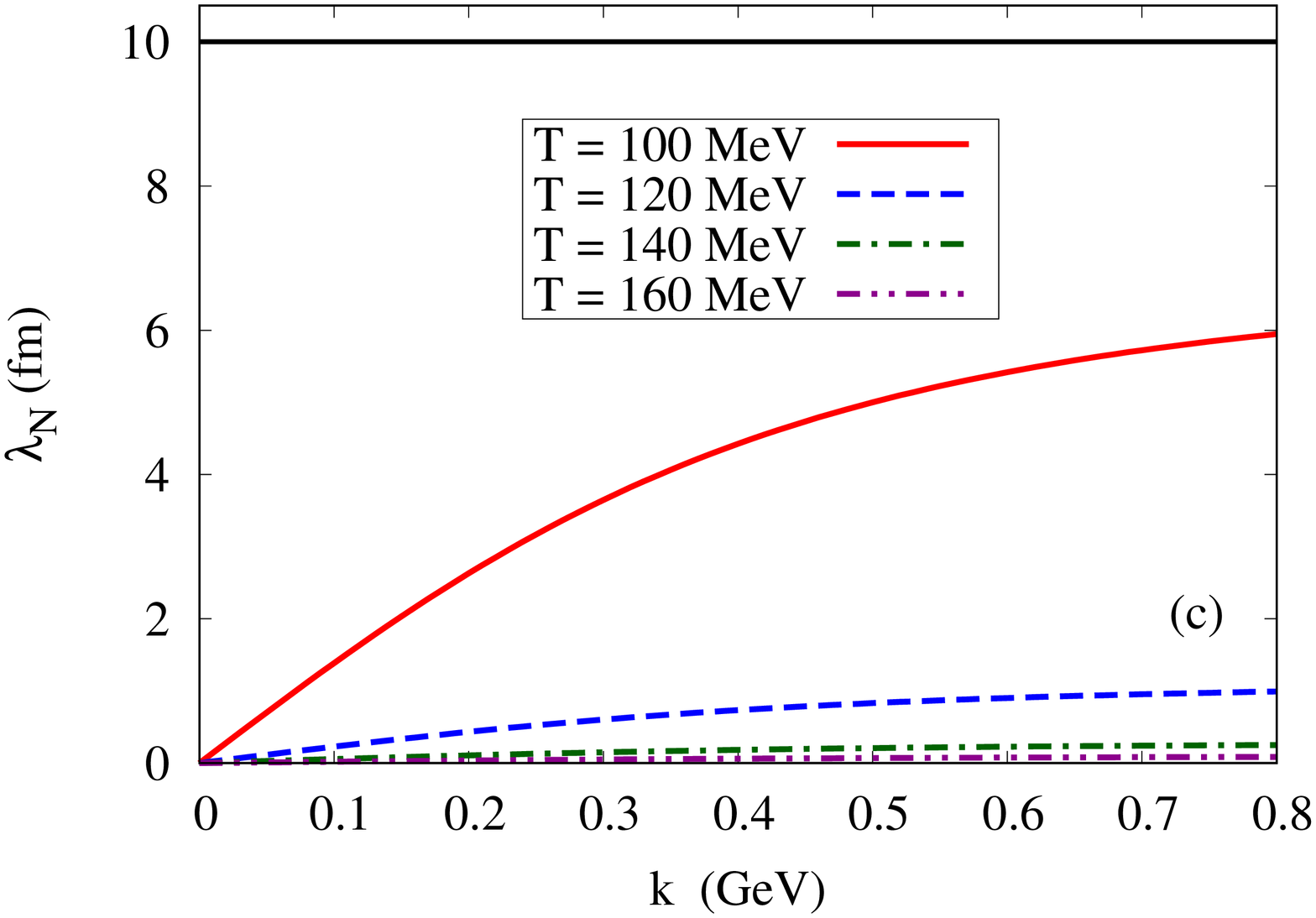}
\caption{(Color online) Momentum distribution of relaxation lengths for (a) $\pi$, (b) $K$ and (c) $N$ 
at different temperatures.}
\label{mf3} 
\end{figure}
%
\begin{figure}[h]
\includegraphics[scale=0.3,angle=-90]{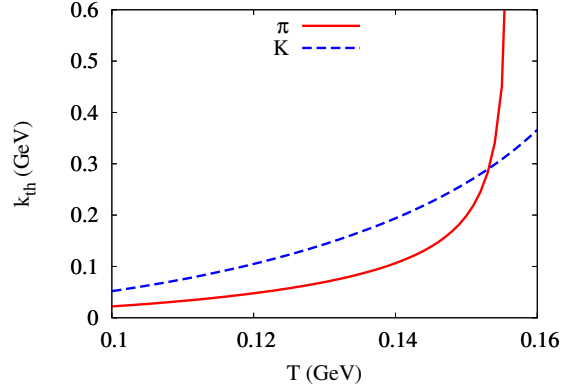}
\caption{(Color online) Temperature dependent of upper momentum threshold $\vk_{\rm th}$ for $\pi$, K.
Beyond the threshold, their relaxation lengths exceed the fireball dimension.}
\label{kth_T} 
\end{figure}

After getting an approximate numerical bands of transport coefficients of hadronic matter,
now let us focus on absolute estimation. If we collect estimated values of transport
coefficients for hadronic matter by earlier studies, then we will get a broad numerical band
within which those estimations are located. In this regards, the present investigation provide a
little narrow band and we are expecting that the values of transport coefficients for hadronic
matter should be located within our proposed band, when one properly take care about the finite 
size dissipation phenomena.
By taking an example, let us demonstrate how to consider
the dissipation of hadrons within the finite size hadronic matter
and how it will help to reshape the values of transport coefficients within
our proposed band.
Let us calculate the relaxation times of NR particles
from the experimentally available data of their
scattering lengths. 
Here, we are considering scattering lengths $R_{ab}^I$ for different isospin ($I$) states
of $\pi\pi$, $\pi N$, $NN$, $KN$ interactions from Refs.~\cite{exp1,exp2,exp3} 
and $\pi K$ interaction from Refs.~\cite{exp4,exp5} and then using these values,
we have calculated isospin average cross sections 
\be
\sigma_{ab}=\sum_{I} (2I+1)4\pi \left|R_{ab}^I\right|^2 \Big/ \sum_{I} (2I+1)~.
\ee
These input details are displayed in Table~(\ref{tab1})~. Now, using these 
isospin average cross sections,
we can calculate the relaxation time $\tau_a$ (a=$\pi$, $K$ and $N$) from the relation,
\be
\frac{1}{\tau_{a}(\vk_a)}=\sum_{b\in\{\pi,K,N\}}\int \frac{d^3\vk_b}{(2\pi)^3}
[\sigma_{ab}v_{ab}~n_{b}]~,
\ee
where $n_b$ is BE/FD distribution function of meson/baryon;
\be
v_{ab}=\left(\frac{1}{2\om_a\om_b}\right)\sqrt{\left\{\frac{}{}s-(m_a +m_b)^2\right\}\left\{\frac{}{}s-(m_a - m_b)^2\right\}}~
\ee
is the relative velocity with $\om_{a,b}=\sqrt{\vk_{a,b}^2+m_{a,b}^2}$ and $s=(\om_{a}+\om_{b})$.
With help of the relaxation time $\tau_{a}$ of $\pi$, $K$ and $N$, one can calculate their
relaxation length $\lambda_a=\vk_a\tau_a/\om_a$, as shown in Fig.~(\ref{mf3}).
Here we see that relaxation lengths
for $\pi$ and $K$ exceed the dimension of fireball in the high momentum domain; although
the nucleon relaxation length always remains lower than the dimension of fireball 
in the entire momentum range. Now, let us use these entire momentum distribution
of $\pi$, $K$ and $N$ in Eqs.~(\ref{eta_G}), (\ref{zeta_G}) to get NR contribution
of $\eta$ and $\zeta$. Then after adding the contribution of R 
component, we will get the total as shown by green dash dot line in 
Figs.~\ref{fig:eta}(a) and \ref{fig:zeta}(a).
Their dimensionless, normalized values, quantified as $\eta/s$ and $\zeta/s$
respectively, are shown by green dash dot line in 
Figs.~\ref{fig:eta}(b) and \ref{fig:zeta}(b).
All curves are going beyond the upper bound of our proposed numerical band.
The reason is that we are
considering high momentum $\pi$ and $K$, which are not at all participants
in dissipation process as their relaxation lengths exceed the system size.
This can be well visualized from Figs.~\ref{mf3}.

To resolve it, we have first track numerically the upper
momentum threshold or cut-off at different temperatures for $\pi$ and $K$, within which their
relaxation lengths don't exceed the fireball dimension (10 fm).
From Fig.~\ref{mf3}, one can visualize this fact graphically
and then we have plotted the the upper momentum thresholds $\vk_{\rm th}$ 
for $\pi$ and K as a function of temperature, which is shown in Fig.~(\ref{kth_T}).
Now when we put those $T$ dependent
momentum thresholds as upper limit in integration of Eq.~(\ref{eta_G}) 
and use the modified results of $\pi$ and $K$, the total
values of $\eta$, $\eta/s$, $\zeta$ and $\zeta/s$ will politely remain within the numerical band. 
It is shown by blue dash lines in Figs.~\ref{fig:eta} and \ref{fig:zeta}.
Hence, our investigation states that the values of 
transport coefficients for hadronic matter will be within our proposed numerical band, 
when one will properly impose the finite size dissipation of NR and R components during
the calculation. We are first time addressing this realistic or phenomenological issue,
which should be considered for transport coefficients calculations of hadronic matter,
which is not an infinite in size.

We may further extend our estimations by adopting folding technique, described by
Eq.~(\ref{fold}). Putting $\eta$, $\zeta$ from Eqs.~(\ref{eta_G}), (\ref{zeta_G})
in $\Phi$, we will get their modified results, as shown by red solid lines
in Fig.~\ref{fig:eta}(a) and~\ref{fig:zeta}(a).
We notice that the values of $\eta$ and $\zeta$ becomes lower due to folding effect.
For convenient of reader, the Table~\ref{tab2} is showing 
our different set of input choices, which we have considered.
\begin{table}[htb]
\begin{tabular}{|l|c|c|}
\hline
  & Transport  & Entropy  \\
& coefficients & density \\
\hline
{Lower bound} & {NR} ($\tau_{\rm low}$) & - \\
&  + {R} ($\tau<10 {\rm fm}$) &  \\
\hline
{Upper bound} & {NR} ($\tau=10$ {fm})  & - \\
& + {R} ($\tau<10 {\rm fm}$) &  \\
\hline
{Set-1} & {NR} + {R} ($\tau<10$ {fm}) & {NR} + {R} \\
\hline
{Set-2} & {NR} ($\tau(\vk)<10$ {fm})   & - \\
& + {R} ($\tau<10$ {fm}) &  \\
\hline
{Set-3} & {Set-2 with folding}  & {Set-1 with folding} \\
\hline
\end{tabular}
\label{tab2}
\caption{Different set of inputs for transport coefficients ($\eta,~\zeta$) 
and entropy density (or other thermodynamical quantities like pressure, speed of sound)}
\end{table}
%



\section{Summary and Discussions}
\label{Sum}
In summary, we have pointed out a phenomenological issue of hadron resonance
gas model, which should be seriously considered during the calculation of shear
viscosity for RHIC or LHC matter and the facts are as follows. At first, on the basis
of the dissipation process, we have classified
our HRG members into two categories - Non-resonance members ($\pi$, $K$ and $N$)
and Resonance members (hadrons other than pseudo-scalar meson nonet and baryon octet).
Former members participate in dissipation via strong interaction scattering processes,
where as the contribution from latter members is coming from their strong decay processes.
We consider only strong interaction processes as other interaction (weak or electromagnetic)
processes are meaningless for this scale of the (hadronic) system~\cite{Purnendu}.
Beyond this normal filtering, we have chosen only those strong decays, whose mean life
times are not exceeding the life time of the hadronic medium, which is roughly chosen as
$10$ fm. Now selecting those resonances and using their mean life times as relaxation
times in the expression of shear viscosity, we get some non-zero value of $\eta$,
which always has to be considered as a background value due to resonances in HRG model.
Taking Compton lengths $(1/m_{\pi,K,N})$
as minimum scattering lengths of NR particles ($\pi$, $K$, $N$), we get a lower
limit estimation for NR component, which has to be added with resonance component.
We notice that the lower limit of total $\eta$, normalized by entropy density $s$ is greater
than its quantum lower bound at high temperature range. Owing to this fact, we may conclude
that $\eta/s$ in HRG model never reach its quantum lower bound near the transition temperature
$T_c$ because of unavoidable resonance contribution. 

Similar to lower limit, the upper limit of shear viscosity can be tuned by equating the 
relaxation length of NR particles to the dimension of medium, produced in
heavy ion experiments. The contribution from resonance component is very definite or known
since it is determined from the experimental values of mean life time of their strong decays,
documented in PDG~\cite{PDG}. Hence, only adjustable quantities are relaxation lengths of NR
particles, whose lower and upper possible values basically give
a narrow numerical band in shear viscosity and other transport coefficients of hadronic matter. 

Next, we have taken a particular example, where absolute values of transport coefficients
are obtained. Here, we have estimated absolute values of relaxation lengths for non-resonance 
particles from their scattering length data. From the momentum distribution of their relaxation lengths,
we have found that pion and kaon relaxation lengths exceed from the fireball dimension beyond
some upper values of momentum, which is again different for different temperature. Now when we
take this temperature dependent momentum cut-off as an upper limit of integration, then the 
values of transport coefficients remain within our proposed numerical band. However, when we
take entire momentum distribution, those values don't remain within the band. Through this
example, we want to emphasize the point - the values of transport coefficients for hadronic matter
will be remain within our proposed numerical band, if we impose the idea of {\it finite size dissipation}. 

%
%
%

\section*{Acknowledgment} 
Snigdha Ghosh acknowledges the Center for Nuclear Theory (CNT),
Variable Energy Cyclotron Centre (VECC), funded by Department of
Atomic Energy (DAE) and Indian Institute of Technology Gandhinagar,
funded by Ministry of Human Resource Development (MHRD) for
support. Sabyasachi Ghosh is partially supported from  University Grant Commission (UGC) 
Dr. D. S. Kothari Post Doctoral Fellowship (India) under
grant No. F.4-2/2006 (BSR)/PH/15-16/0060. 
SB is supported from CSIR. We also thank to Rajarshi Ray and Vinod Chandra for their useful
suggestions.

\appendix

\section{HRG} \label{appendix.hrg}

The momentum integral in Eq.~(\ref{eq.partition.function}) can be performed analytically and be expressed 
in terms of modified Bessel functions $K_n(x)$. Bellow we summarize the final forms of all the thermodynamic 
quantities:
\begin{eqnarray}
\lnz &=& \frac{VT^3}{2\pi^2}\sum_{h\in\{\text{hadrons}\}}^{}\sum_{n=1}^{\infty}g_h\frac{(a_h)^{n+1}}{n^2}\mt^2K_2\nmt \exp\nmut \\
P &=& \frac{T^4}{2\pi^2}\sum_{h\in\{\text{hadrons}\}}^{}\sum_{n=1}^{\infty}g_h\frac{(a_h)^{n+1}}{n^2}\mt^2K_2\nmt \exp\nmut \\
\varepsilon &=& \frac{T^4}{2\pi^2}\sum_{h\in\{\text{hadrons}\}}^{}\sum_{n=1}^{\infty}g_h\frac{(a_h)^{n+1}}{n^2}\mt^2\exp\nmut\left[ \left\{1-\nmut\right\}K_2\nmt \right. \nn \\ 
&& \hspace{3.5cm}\left. +\frac{1}{2}\nmt\left\{K_1\nmt +K_3\nmt\right\}\right] \\
n_k &=& \frac{T^3}{2\pi^2}\sum_{h\in\{\text{hadrons}\}}^{}\sum_{n=1}^{\infty}g_hq_h^k\frac{(a_h)^{n+1}}{n}\mt^2K_2\nmt \exp\nmut 
~~;~~ k = B,Q,S,... \\
\left(\frac{\del P}{\del T}\right) &=& \frac{T^3}{2\pi^2}\sum_{h\in\{\text{hadrons}\}}^{}\sum_{n=1}^{\infty}g_h\frac{(a_h)^{n+1}}{n^2}\mt^2\exp\nmut
\left[ \left\{2-\nmut\right\}K_2\nmt \right. \nn \\
&& \hspace{3.5cm} \left. + \frac{1}{2}\nmt\left\{K_1\nmt +K_3\nmt \right\} \right] \\
\left(\frac{\del\varepsilon}{\del T}\right) &=& \frac{T^3}{2\pi^2}\sum_{h\in\{\text{hadrons}\}}^{}\sum_{n=1}^{\infty}g_h\frac{(a_h)^{n+1}}{n^2}\mt^2\exp\nmut
\left[ \frac{1}{4}\nmt^2\left\{K_0\nmt+K_4\nmt\right\} \right. \nn \\ && \hspace{3.5cm} \left. +\frac{1}{2}\nmt\left\{3-\nmut\right\}\left\{K_1\nmt+K_3\nmt\right\} \right. \nn \\ && \hspace{3.5cm} \left.
+\left\{2-2\nmut+\nmut^2+\frac{1}{2}\nmt^2\right\}K_2\nmt\right] \\
\left(\frac{\del P}{\del\mu_k}\right)&=& n_k ~~;~~ k = B,Q,S,.... \\
\left(\frac{\del \varepsilon}{\del\mu_k}\right)&=& \frac{T^3}{2\pi^2}\sum_{h\in\{\text{hadrons}\}}^{}\sum_{n=1}^{\infty}g_hq_h^k\frac{(a_h)^{n+1}}{n}\mt^2\exp\nmut
\left[ \frac{1}{2}\nmt\left\{K_1\nmt+K_3\nmt\right\} \right. \nn \\ 
&& \hspace{3.5cm} \left. -\nmut K_2\nmt\right] ~~;~~ k = B,Q,S,...
\end{eqnarray}


\end{document}